\documentclass[longnamesfirst,sort,authoryear,round]{interact}

\usepackage{epstopdf}
\usepackage[caption=false]{subfig}

\usepackage[longnamesfirst,sort]{natbib}

\usepackage{array}      
\usepackage{booktabs}   
\usepackage{float}
 \usepackage{graphicx}
 \usepackage{subcaption}
\usepackage{subfig}
\usepackage{caption}
\usepackage{float} 
\usepackage{natbib}
\usepackage[colorlinks=true]{hyperref}
\usepackage{hyperref}
\hypersetup{
    colorlinks=true,
    breaklinks=true,
    plainpages=false,
    citecolor=black,
    linkcolor=black,
    urlcolor=black,
    bookmarksopen=true,
    bookmarksnumbered=false,
    bookmarksdepth=5
}
\usepackage{natbib}

\usepackage[dvipsnames]{xcolor}
\usepackage{longtable}
\usepackage{enumitem}
\usepackage{amsmath,amssymb}

\usepackage{comment}

\bibpunct[, ]{(}{)}{;}{a}{,}{,}
\renewcommand\bibfont{\fontsize{10}{12}\selectfont}

\theoremstyle{plain}

\theoremstyle{definition}

\theoremstyle{remark}

\begin{document}


\title{Context-Aware Optimization of Follow-Up Intervals for Type 2 Diabetes Care Using Markov Decision Processes}

\author{
\name{Parisa Lotfibagha\textsuperscript{a}, Kristen Miller\textsuperscript{b, c, d}, William J. Gallagher\textsuperscript{b, d}, Elizabeth B. Selden\textsuperscript{e}, \thanks{CONTACT M.~C. Author. Email: mcapan@umass.edu}Muge Capan\textsuperscript{a}}
\affil{\textsuperscript{a}University of Massachusetts Amherst, United States, \textsuperscript{b}MedStar Health Center for Diagnostic Systems Safety, United States, \textsuperscript{c}National Center for Human Factors in Healthcare, MedStar Health Research Institute, United States, \textsuperscript{d}Georgetown University School of Medicine, United States,
\textsuperscript{e}Medstar Georgetown University Hospital, United States
}    
}
\maketitle


\begin{abstract}
Chronic disease management relies on regular patient-provider interactions to follow-up on disease progression and control. For Type 2 Diabetes (T2D), current guidelines prescribe fixed time intervals between subsequent primary care visits for all patients, overlooking heterogeneity in clinical trajectories and patient characteristics. This study introduces a Contextual Markov Decision Process (CMDP) model to optimize subpopulation-specific follow-up interval decisions using Electronic Health Record (EHR) data from 22,154 T2D patients across 10 primary care clinics. Contexts are identified by: i) dimensionality reduction of variables representing the individual health trajectories utilizing Principal Component Analysis, and ii) assigning patients to contexts via principal components and additional patient-level features using clustering. Two distinct contexts emerged, representing a lower- and a higher-risk subpopulation. CMDP-derived policies recommend: (i) follow-up within 1 month if lab value at current visit is unmeasured; (ii) up to 3 months for elevated lab values or recent hospitalizations; and (iii) 6 to 12 months for sustained glycemic control, with shorter follow-up intervals for patients in high-risk context. The optimal policies achieved lower expected cumulative cost than benchmarks (e.g., in the higher-comorbidity context, the CMDP policy reduced cost by about 34.8\%, and in the lower-comorbidity context by about 6.4\%, relative to an American Diabetes Association-like fixed interval follow-up policy. These findings demonstrate how context-aware approaches can inform adaptive follow-up strategies, and have the potential to advance chronic care management in primary care by synthesizing machine learning and probabilistic decision models.
\end{abstract}



\section{Introduction}
The increasing prevalence of chronic diseases is pressuring healthcare systems to design adaptive follow-up strategies that reflect patient heterogeneity and lifelong care requirements. Type 2 Diabetes (T2D), affecting over 589 million adults globally, exemplifies this challenge \citep{ceriello2025idf}. Regular post-diagnosis follow-up is essential to maintain disease control and prevent further T2D-induced health complications \citep{american2025introduction}. In the United States, post-diagnosis follow-up of T2D commonly occurs during visits with primary care physicians (PCPs) to achieve and maintain \textit{glycemic control} \citep{kushner2022role}. Glycemic control is measured by a biomarker -- hemoglobin A1c (HbA1c) -- which represents the average blood sugar level over the past three months. The American Diabetes Association (ADA) guideline defines glycemic control using the target value of HbA1c $\le$7\% (53 mmol/mol) \citep{american20256}. 

Without regular follow-up, majority of adult T2D patients struggle to achieve and maintain this HbA1c target \citep{yahaya2023poor, njonnou2024factors}. ADA guidelines recommend quarterly PCP follow-up for T2D patients whose therapy has changed or who have not achieved the HbA1c target, and at least twice yearly for those who have achieved glycemic control \citep{elsayed20236}. Finding the optimal and personalized follow-up interval between subsequent PCP visits is a complex decision problem. This decision requires a delicate balance. If the time between subsequent PCP visits is too short, this can increase the frequency of visits and overwhelm patients as well as the healthcare system. On the other hand, long follow-up intervals increase the risk of failing to detect lack of glycemic control \citep{yahaya2023poor}. 

Several studies explored the problem of identifying a risk-appropriate follow-up interval for individuals diagnosed with T2D. For example, a randomized trial in adults with T2D who exhibit HbA1c values consistently $\le$7.5\% found that monitoring every three and every six months yielded comparable health outcomes \citep{wermeling2014effectiveness}. Another study found that regular follow-up interval of two months maintained glycemic control at rates comparable to monthly follow-up \citep{ukai2019effectiveness}. For patients who failed to achieve glycemic control, a cohort study demonstrated that having more than 4 PCP visits per year increased the likelihood of achieving glycemic target (HbA1c $\le$ 9.5\% [80  mmol/mol]) compared to lower frequency follow-up groups \citep{asao2014revisit}. The positive impact of increased follow-up frequency on achieving glycemic control was confirmed by another cohort study, which found that more frequent encounters were associated with faster attainment of an HbA1c target of $<$7\% (53 mmol/mol)--by about 35\% in non-insulin-treated patients and 17\% in insulin-treated patients \citep{morrison2011encounter}. However, these studies focus on associations between fixed visit intervals (e.g., 3 vs. 6 months) and outcomes, typically analyzed using descriptive or regression-based methods that estimate population-average effects, and lack a prescriptive framework for optimizing visit intervals considering patients' evolving clinical trajectories.

In this study, we approach finding the optimal post-diagnosis follow-up interval for T2D patients as a dynamic decision problem under uncertainty. Specifically, we introduce a Contextual Markov Decision Process (CMDP) model that uses longitudinal EHR data, and integrates dimensionality reduction and clustering to learn context-specific follow-up policies. Markov Decision Process (MDP) models have been widely used in chronic disease screening and treatment decisions \citep{ayer2012or, hajjar2023personalized}. Specifically in T2D, studies compared MDP-derived medication treatment strategies to local clinical practice. Results showed that MDP policy led to up to 0.27 additional QALYs for high-risk patients (e.g., age at diagnosis of 60, smokers, or with HbA1c levels of  9\% and above), whereas the improvements for low-risk patients were negligible \citep{meng2020analysis}. Focusing on the screening of undiagnosed T2D populations aged 45 and older, \cite{wu2022optimizing} formulated a partially observable MDP to determine the optimal frequency of HbA1c testing. Their model recommended less frequent screening (e.g., every 3-6 years) for lower-BMI individuals aged 45-65, but significantly more frequent screening (e.g., every 1-3 years) for obese adults (BMI $\ge$30) aged 65 and older. Despite these advances, existing MDP models for T2D have largely overlooked the problem of optimizing post-diagnosis follow-up intervals in PCP. Prior studies have focused on treatment selection or one-time screening frequency and estimated policies for broad risk groups. Thus, there is a need to capture the dynamic clinical trajectories inherent and provide context-specific follow-up recommendations for T2D patients.

To address this gap, we formulate a CMDP to explicitly model trade-offs between time spent in uncontrolled glycemic health states and healthcare utilization, and inform context-aware follow-up for individuals diagnosed with T2D. Clinical contributions of this study include the conceptualization of trade-offs between time spent in uncontrolled glycemic states and hospitalizations between subsequent PCP visits using Electronic Health Records (EHR) data and operationalization of follow-up interval decisions in PCP. The methodological novelty relies on the application of dimensionality reduction on complex post-diagnosis clinical trajectories and mixed-data clustering methods, and their synthesis within a Markov decision modeling framework to inform context-aware follow-up decision making in primary care setting.


\section{Materials and Methods}\label{sec:meterial-methods}
\subsection{Study Population}\label{sec:population}
This is a retrospective observational cohort study of patients who received care at one of the ten primary care clinics within an integrated healthcare system in the mid-Atlantic region of the United States. The study period spanned January 1, 2017, through December 31, 2023. The inclusion criteria were: (i) age 18 years or older at the time of their initial primary care encounter; (ii) a documented diagnosis of T2D, as identified by relevant International Classification of Diseases, Tenth Revision (ICD-10) codes (Appendix~\ref{apendixA}) \citep{WHO2022}; and (iii) at least three PCP visits within the study window. The sample size was 
20,483 unique patients and 1,399,615 unique visits. The research protocol was approved by an Institutional Review Board (IRB), and the health system authorized the use of de-identified EHR data.

\subsection{Methods}\label{sec:methods}
\begin{figure}[H]
  \centering
   \makebox[\linewidth][c]{%
      \includegraphics[width=1.1\linewidth]{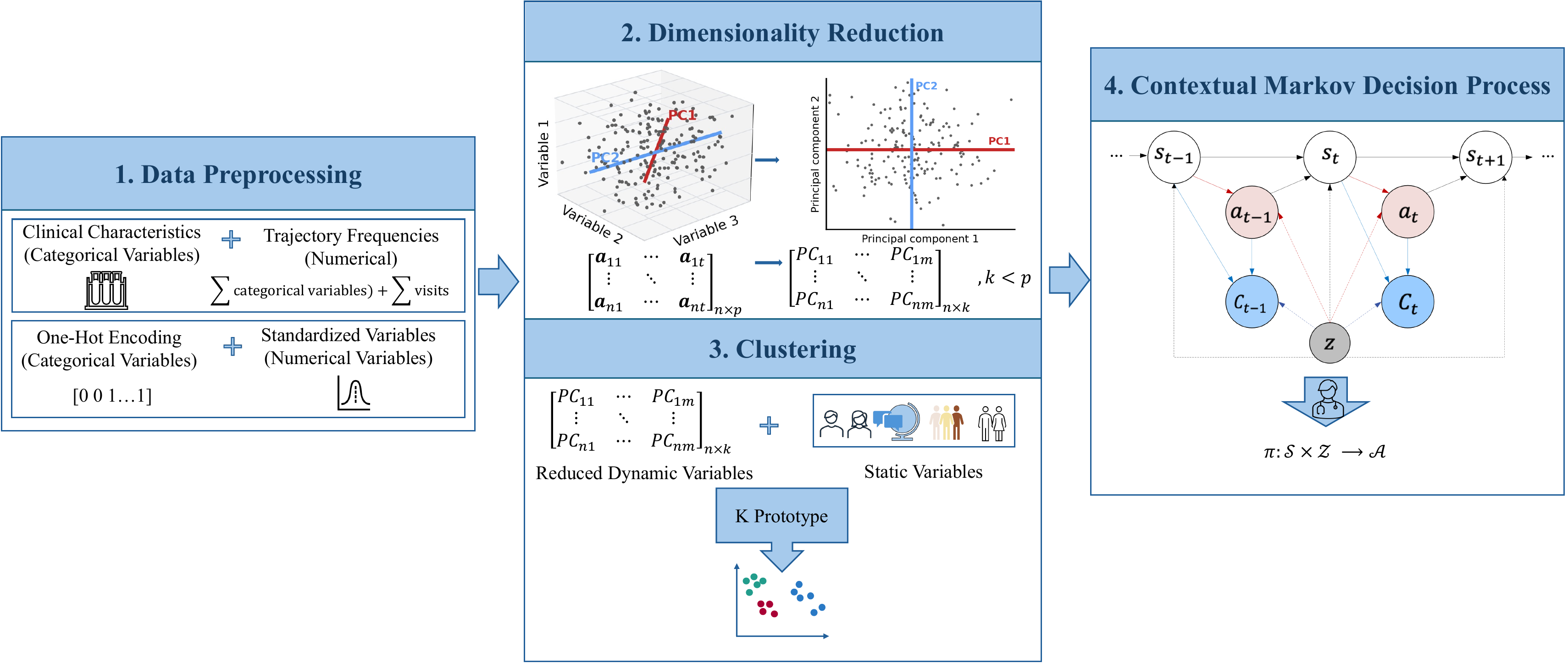}
      }
  \caption{Overview of the methodological approach for deriving context-specific follow-up policies. PC stands for principal component. $n$ represent the number of unique patients in study population. Dynamic variables are observed at each PCP visit at time $t$. After PCA, the dimension is reduced to $k$. $s_t$ represents health state at PCP visit $t$, $a_t$ the follow-up action, $c_t$ cost accumulated between two subsequent PCP visits at times $t_1$ and $t$, and $z$ stands for the context.}
  \label{fig:pipeline}
\end{figure}

Our analytical framework, illustrated in Figure~\ref{fig:pipeline}, is a multi-stage approach to learn context-specific follow-up policies from EHR data. The framework consists of four main stages: (i) Data Preprocessing, where raw static and dynamic clinical data are transformed into standardized variable sets; (ii) Dimensionality Reduction, where high-dimensional patient trajectories compressed into dense embeddings; (iii) Clustering, where patients are clustered into different contexts; and (iv) Policy Optimization, where a CMDP is solved to derive optimal follow-up intervals for each subpopulation.


\subsubsection*{ Data Preprocessing} \label{sec:inputdata}
The first stage involves preprocessing raw EHR observations into a standardized patient-level variable set comprising both static and dynamic variables. The \textit{static variables} refer to data elements observed for each patient at their index PCP visit and do not change over time. These include: (i) demographics (e.g., sex, age at diagnosis); (ii) clinical characteristics (e.g., presence of comorbid conditions such as cardiovascular disease (CVD), chronic kidney disease (CKD), and T2D complication status indicating health complications induced by T2D, defined using appropriate ICD-10 codes as detailed in Appendix~\ref{A}); and (iii) healthcare utilization indices (e.g., total number of hospitalizations and total number of PCP visits over study period). The selection of these static variables is guided by evidence that demographic characteristics, comorbidities, and patterns of care utilization are key determinants of glycemic control, and adverse outcomes. Age at diagnosis and sex have been associated with differences in glycemic control and complication risk among individuals with T2D \citep{mahmood2016glycaemic}, while comorbid conditions such as CVD and CKD have been linked to poor glycemic control and high complication burden \citep{bitew2023prevalence}. 

The \textit{dynamic variables} refer to time-dependent observations at each PCP visit ($t$) for each patient throughout the study period. These variables characterize current glycemic control, short-term changes in glycemic level, and recent acute events. 
\begin{itemize}
    \item \textit{Glycemic status at time t:} The patient's HbA1c level, discretized into three categories based on the most recent measurement at or before visit $t$: \textit{In-Control} (HbA1c $<$6.5\% [48 mmol/mol]), \textit{Out-of-Control} (HbA1c $\ge$6.5\% [48 mmol/mol]), or \textit{Unmeasured} if no prior HbA1c is on record. The threshold of 6.5\% (48 mmol/mol) is consistent with American Association of Clinical Endocrinologists clinical guideline that recommends an HbA1c target of 6.5\% (48 mmol/mol) for patients with T2D \citep{garber2020consensus}.
    \item \textit{Hospitalizations between subsequent PCP visits:} A binary indicator of any inpatient admission between two consecutive PCP visits (1 = had at least one or more hospitalizations between subsequent PCP visits, 0 = otherwise). Interval hospitalization serve as markers of acute clinical deterioration and multimorbidity burden and are associated with subsequent adverse outcomes and higher healthcare utilization \citep{schneider2016diabetes, khalid2014rates}, making them relevant for decision about intensifying follow-up \citep{american202516}.
    \item \textit{HbA1c trend between subsequent PCP visits:} A categorized variable capturing the change \(\Delta\text{HbA1c} = \text{HbA1c}_t - \text{HbA1c}_{t-1}\) from PCP visit at time $t-1$ to PCP visit at time $t$, categorized into the following categories:  
    \begin{itemize}
        \item Improving: \(\Delta\text{HbA1c} <-0.5\%\),
         \item Stable: \(|\Delta\text{HbA1c}| \le 0.5\%\),
        \item Worsening: \(\Delta\text{HbA1c} >0.5\%\), and .
       \item Undefined: If glycemic status was unmeasured at PCP visit at either time $t$, time $t-1$, or both.
    \end{itemize} 
    The reason for including short-term HbA1c trend is because visit-to-visit glucose fluctuations have been linked diabetes-related health complications \citep{suh2015glycemic, zhou2020glycemic}. Evidence shows that reducing short-term glycemic fluctuations can improve the severity of uncontrolled T2D \citep{monnier2018application}. We use a cut-off of 0.5\% to define improvement and worsening because HbA1c changes of this magnitude are commonly interpreted as clinically meaningful differences in studies of glycemic outcomes \citep{chen2022long}
\end{itemize}

\subsubsection*{Principal Component Analysis}\label{sec:methods-context}
The second stage in the analytical approach is dimensionality reduction (Figure~\ref{fig:pipeline}). In this study, the high dimensionality of the clinical trajectories arises from representing each patient’s sequence of clinical states across PCP visits over the study period. At each PCP visit, the clinical state is defined as the Cartesian combination of glycemic status, HbA1c trend, and interval hospitalization. With 3 glycemic categories, 4 trend categories, and 2 hospitalization categories, this results in 24 distinct clinical trajectory states (\(3\times 4\times 2=24\)) that a patient can be in at a given visit $t \in \{1, \cdots,T\}$. All trajectories are normalized to a uniform length \(T_{\max}=16\), chosen to cover 95\% of visit counts in the cohort; for longer sequences, we retain the most recent \(T_{\max}\) visits, and shorter sequences are right-padded with a Loss-to-Follow-Up (LTF) token. 


Let $X = (x_{ij})$ denote the resulting feature matrix, with rows representing unique patients $i = 1,\dots,n$ and columns representing variables preprocessed in previous stage $j = 1,\dots,p$. We use principal component analysis (PCA) to reduce the dimensionality of $X$ while transforming correlated variables into a smaller set of uncorrelated components that retain the most of variation in the original dynamic variables \citep{boehmke2019hands}. We determined the number of principal components (PCs) using the cumulative proportion of variance explained. After computing the eigenvalues of the variable covariance matrix, we examined the cumulative proportion of variance explained by successive PCs and retained the smallest number of components that together accounted for at least 95\% of the total variance. This threshold balances dimensionality reduction with preservation of the information contained in the original trajectory features.

\subsubsection*{Clustering}\label{sec:context-kproto}
The goal of the clustering stage is to assign patients to contexts. Using the 83 selected PCs together with 5 static variables (sex, age at diagnosis, presence of comorbid conditions such as CVD, CKD, and T2D complication status) as input, we identified $K$ patient contexts with the K-Prototypes algorithm, a mixed-type clustering method designed to handle numeric and categorical variables \citep{huang1998extensions, ahmad2007k}. We treated the PCs as numeric variables and the static patient variables as categorical variables. Prior to clustering, the PCs for each patient were standardized to a zero mean and unit variance. Standardization is important for distance-based clustering methods such as K-prototype to ensure that all variables contribute comparably to the distance calculation and prevents those with larger ranges from dominating the results \citep{hastie2009elements}.
The K-Prototype algorithm, which is used for mixed numeric and categorical data, constructs clusters by minimizing the total within-cluster distance by using the mean for numerical variables and the mode for categorical variables \citep{huang1998extensions, preud2021head, pasin2023investigation}. We selected the number of clusters $K$ that maximized the mean silhouette score. The final model assigned each patient a context label $z\in\{1,\ldots,k\}$ for downstream use in the CMDP transition and cost models.

\subsubsection*{Contextual Markov Decision Process (CMDP)} \label{sec:cmdp}
In the last stage of the methodological approach, we model finding the optimal return-to-clinic interval as an infinite-horizon discounted \textit{contextual Markov decision process}. Each decision epoch corresponds to a PCP visit at which the follow-up interval is selected. The CMDP is defined by the tuple $(\mathcal{S},\mathcal{A},\{P_z\},\{C_z\},\gamma)$, where $\mathcal{S}$  denotes the clinical state space, $\mathcal{A}$ is the set of feasible follow-up intervals,$P_z$ is the context-specific state transition, $C_z$ is the immediate context-specific cost function that aggregates clinically interpretable components, and $\gamma \in (0,1)$ is the discount factor. Each patient is assigned a context $z \in \{1,\ldots,K\}$, with context-specific state transition probabilities $P_z(s_{t+1} \mid s_t,a)$ and immediate cost function $C_z(s_t,a,s_{t+1})$.  For each context $z$, our objective is to compute an optimal policy $\pi_z:\mathcal{S}\to\mathcal{A}$ that minimizes expected discounted cumulative cost. The components of the CMDP are described in detail below.

\paragraph*{\textit{State Space.}} 
The state at each decision epoch is defined as the patient’s clinical status. Specifically, at decision epoch $t$ (a completed primary care visit), the patient's clinical status is 
\begin{align}
s_t =\big(s_t^H,\; s_t^I,\; s^{\Delta H}_t\big),
\end{align}
where the components are categorical variables defined in Section~\ref{sec:inputdata}: $s_t^H \in \mathcal{S}^H=\{\text{In-Control},\text{Out-of-Control},\text{Unmeasured}\}$(glycemic status),
$s_t^I \in \mathcal{S}^I=\{\text{No},\text{Yes}\}$ (interval hospitalization),
and $s^{\Delta H}_t \in \mathcal{S}^{\Delta H}=\{\text{Improving},\text{Stable},\text{Worsening},\text{Undefined}\}$ (short-horizon HbA1c trend).

The full state space, $\mathcal{S}$, is defined as the union of the Cartesian product of these non-absorbing component spaces and a  terminal absorbing state:
\begin{align}
\mathcal{S}= \big(\mathcal{S}^H \times \mathcal{S}^I \times \mathcal{S}^{\Delta H}\big)\, \cup \,\{\text{LTF}\},    
\end{align}
with $\big|\mathcal{S}^H \times \mathcal{S}^I \times \mathcal{S}^{\Delta H}\big| = 3 \times 2 \times 4 = 24$. The absorbing state, \textit{LTF} (Loss-to-Follow-Up), is entered when a patient has no subsequent PCP visit within the study observation window. Upon entering this state, the decision process terminates.
\paragraph*{\textit{Action Space.}} At decision epoch $t$, given state $s_t$, the action $a_t\in\mathcal{A}$ denotes the return-to-clinic interval for the next PCP visit. Because explicit provider follow-up prescriptions are not recorded in the EHR, we used the realized time to the next PCP visit as a proxy for the historical action during model estimation. This continuous interval is discretized into five month-based categories (M = month), defining 
\begin{align}
 \mathcal{A}=\{\text{1M},\,\text{3M},\,\text{6M},\,\text{12M},\,\text{$>$12M}\}.   
\end{align}

For each patient, the final PCP visit observed during the study period (i.e., when no subsequent PCP visit is recorded) is represented by a terminal absorbing state and assigning the longest allowable follow-up action $(a_t= >$12M).

\paragraph*{\textit{Cost Structure.}} 
Costs are non-monetary and quantify the clinical burden accumulated in each health state and each action. Total immediate cost collected between two subsequent PCP visits is the sum of four components: (i) time spent in uncontrolled glycemic state represented as the integrated area above the HbA1c target level ($\geq$ 6.5\%) multiplied with the follow-up time interval, (ii) measurement uncertainty due to delayed or unavailable HbA1c results, (iii) the burden from interval inpatient admissions, and (iv) visit utilization burden reflecting the frequency of primary care follow-up. Each component is described below.
\begin{itemize}
\item  \textit{Time and Severity of Uncontrolled Glycemic State} ($C^\mathrm{G}$): This component measures cumulative exposure to hyperglycemia above the clinical threshold $\tau = 6.5\%\,(48~\text{mmol/mol})$ over the inter-visit interval of length $a$ (months). We integrate the area above the target HbA1c level over the follow-up interval to represent the glycemic burden accumulated between two subsequent PCP visits and scale this quantity by a state-transition adversity weight $W(s_t,s_{t+1})$ that differentially weights the penalty according to the type of transition (e.g., assigning higher weights to transitions that worsen or increase uncertainty about glycemic control, such as In-Control $\rightarrow$ Out-of-Control, and lower weights to transitions that improve control, such as Out-of-Control $\rightarrow$ In-Control). Assuming a linear evolution of HbA1c between visits, we approximate the weighted integral using the trapezoidal rule as 
\begin{align}
  C^{\mathrm{G}}(s_t,a_t,s_{t+1}) = W(s_t,s_{t+1}) \cdot \Bigl[\tfrac{1}{2}\,(E_{s}+E_{s_{t+1}}) \cdot a\Bigr],
\end{align}
    
where $E_{s_t} = \max\{0,\mathrm{HbA1c}(s_t) - \tau\}$ and $E_{s_{t+1}} = \max\{0,\mathrm{HbA1c}(s_{t+1}) - \tau\}$  denote exceedance above target at the beginning (state $s$) and end (state $s_{t+1}$) of the interval. Here, $\mathrm{HbA1c}(s_t)$ (and analogously $\mathrm{HbA1c}(s_{t+1})$) is the numeric value associated with the state: when a measurement exists at or before the index visit, we use last-observation-carried-forward (LOCF); otherwise, we use a pre-specified belief value $\kappa$ (and $\kappa_{\mathrm{LTF}}$ if the transition terminates in LTF). The weight $W(s_t,s_{t+1})$ is dimensionless and reflects the relative adversity of the transition, with larger values assigned to clinically undesirable transitions and smaller values to clinically favorable transitions.

\item \textit{Cost of Unobserved HbA1c} ($C^\mathrm{M}$): This component penalizes follow-up decisions made when the patient's current HbA1c level is unobserved or based on outdated measurements. The penalty increases both with the degree of uncertainty about HbA1c and with the length of the selected follow-up interval. We define a nonnegative \emph{uncertainty level} $u_t \ge 0$ that summarizes how informative the available HbA1c information is at decision epoch $t$:
    \begin{itemize}
        \item if $s^H_t \in \{\text{In-Control},\text{Out-of-Control}\}$, $u_t$ is derived from the time since the last observed HbA1c measurement;
        \item if $s^H_t  = \text{Unmeasured}$ and this is not the first PCP visit, $u_t$ is derived from the length of the previously chosen follow-up interval;
        \item if $s^H_t  = \text{Unmeasured}$ at the first PCP visit, there is no prior measurement or action, and we instead apply a separate baseline penalty.
    \end{itemize}

    Let $d_t$ denote the number of days since the last available HbA1c at epoch $t$. When a measurement exists, $d_t$ is defined as the time difference between the current visit and the last HbA1c result. When no prior HbA1c measurement is available but $t>1$ and $s_t^H = \text{Unmeasured}$, $d_t$ is approximated using the length of the previous follow-up interval. We then map $d_t$ to an ordinal uncertainty level via a nondecreasing step function,
    \[
    u_t \;=\; \lambda_{\mathrm{u}}(d_t) \;=\;
    \begin{cases}
    0,  & d_t \le 45,\\
    1,  & 45 < d_t \le 90,\\
    2,  & 90 < d_t \le 180,\\
    4,  & 180 < d_t \le 366,\\
    16, & d_t > 366.
    \end{cases}
    \]
    The resulting uncertainty cost at epoch $t$ is
    \begin{align}
    C^{\mathrm{M}}(s_t,a_t,s_{t+1})
    =
    \begin{cases}
    \eta\, a_t, & s^H_t = \text{Unmeasured} \text{ and $t$ is the first PCP visit},\\[4pt]
    u_t \, a_t, & \text{otherwise},
    \end{cases}
    \end{align}
    
    where $\eta \ge 0$ is a scaling parameter and the impact of varying $\eta$ is examined in sensitivity analyses (Section~\ref{sec:sensitivity}).

    In this formulation, longer follow-up intervals are penalized more heavily when HbA1c information is less recent or unavailable, reflecting the increased clinical risk associated with prolonged periods of glycemic uncertainty.
  
  \item \textit{Cost of Hospitalization} ($C^{\mathrm{I}}$): Between consecutive PCP visits, a patient may have hospital admissions. Let
    \[H(s_{t+1}) = \mathbf{1}\!\{\, s_I^{t+1} = \text{Yes} \,\}\]
    indicate that at least one admission occurred in the interval \((t,\,t{+}1]\) (as recorded at visit \(t+1\)); otherwise \(H(s_{t+1})=0\).
    We penalize any such interval by
    \begin{align}
        C^{\mathrm{I}}(s_t,a_t,s_{t+1}) = \lambda_{\mathrm{I}}\, H(s_{t+1}),
    \end{align}
    where \(\lambda_{\mathrm{I}}>0\) is a scaling parameter that reflects the relative burden associated with hospitalization between PCP visits.
 \item  \textit{Premature Return to PCP Cost} ($C^{\mathrm{F}}$): Considering that HbA1c reflects average glycemic control over approximately 3 months \citep{elsayed20236}, very short very short return intervals to PCP may not add any additional information value on disease trajectory. On the contrary, it may impose unnecessary burden on patients and the delivery system due to increased frequency of PCP visits. This cost element reflects the required balance between following up too soon (and not obtaining new information) vs. too late (and missing the opportunity to observe elevated HbA1c values and intervene) when deciding the length of follow-up time interval between subsequent PCP visits.
    
To reflect this, we include a utilization cost that depends only on the action $a_t$ in months. We choose a concave, decreasing function of $a_t$ via a transformed logarithm,
    \begin{align}
        C^{F} (s_t,a_t,s_{t+1}) = \lambda_{F} [\log (\frac{20}{a_t})]^{1.5},
    \end{align}

 where $\lambda_{F} > 0$ is a scaling parameter, and the constant 20 (months) sets a reference horizon beyond the longest action in $\mathcal{A}$. This term strongly penalizes extremely frequent follow-up (e.g., 1 month) relative to more routine intervals, while assigning negligible cost to the longest follow-up option.
\end{itemize}

\section{Results}
\subsection{Descriptive Statistics of Study Population}
The final analytic cohort included 20,483 unique patients. The cohort was predominantly female (59.94\%) and African American (66.55\%), with a majority (91.90\%) identifying as Non-Hispanic. The patient population was largely middle-aged and older, with 52.59\% of patients (n=10,772) being 60 years or older. Nearly half of all patients (47.45\%) had a documented T2D-related complication, 34.45\% had CVD, and 25.77\% had CKD. Table~\ref{tab:characteristics} summarizes the study population characteristics.

\subsection{PCA Results} \label{sec:results-pca}
\begin{figure}[H]
  \centering
  \includegraphics[width=0.9\textwidth]{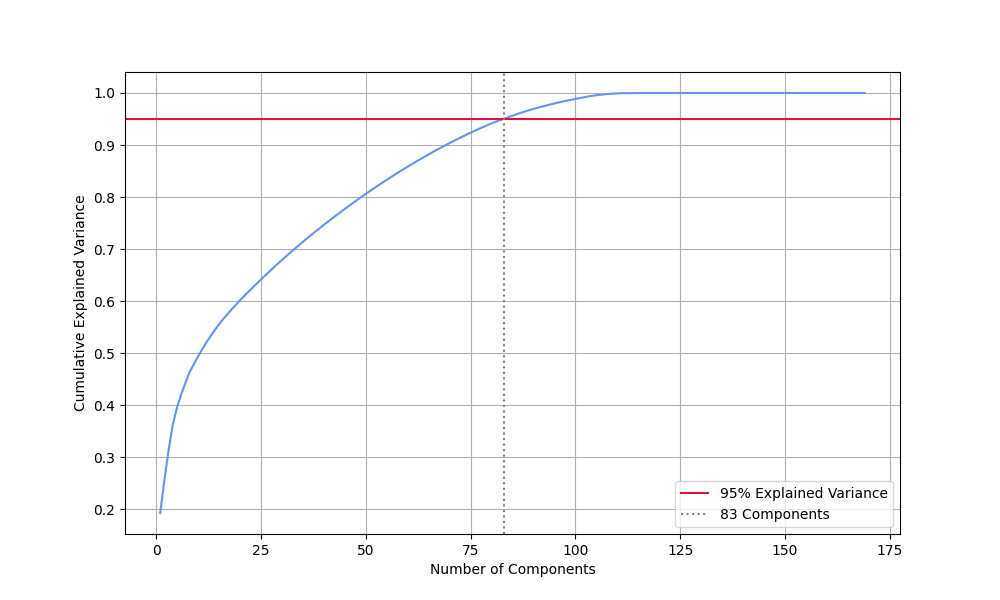}
  \caption{Selection of principal components based on cumulative explained variance. The plot shows the cumulative proportion of variance (y-axis) explained by the principal components (x-axis). The dashed vertical line indicates that 83 components were required to capture the pre-specified 95\% (red horizontal line) of the total variance in the standardized feature set.}
  \label{fig:pca_component}
\end{figure}

Figure~\ref{fig:pca_component} plots the cumulative explained variance as a function of the number of principal components.  Figure~\ref{fig:pca_component} illustrates a sharp initial increase followed by a plateau, indicating that each additional PC explains progressively less variance. As shown by the intersection with our 95\% threshold, 83 PCs were retained from the original 409-dimensional variable space (400-dimensional trajectory sequence and 9 aggregate variables). 


\subsection{Clustering Results}\label{sec:results-contexts}

The optimal number of clusters, $K$, for the K-Prototypes algorithm were selected based on the mean silhouette score computed using Gower dissimilarity matrix. Results showed that the silhouette score was maximized at $K=2$ (0.252). The score dropped substantially for $K=3$ (0.103), $k=4$ (0.032), and remained low at $K=5$ (0.104). We therefore proceeded with $K=2$ representing two distinct contexts for developing the CMPD model. The comparison of these two contexts with regards to demographic and clinical characteristics is shown in Table~\ref{tab:cluster_outcome}.

Based on clinical domain expert insights (W.G. and E.S.), these two contexts represent distinct patient subpopulations with different follow-up needs as observed in primary care practice. Specifically, context (cluster) 1 represents an older, higher-risk cohort. Patients in this context were predominantly 60 years or older (64.1\%), with a substantially higher prevalence of comorbid CKD (34.45\% vs. 23.96\%), CVD (40.69\% vs. 33.15\%), and existing T2D-related complications (57.95\% vs. 45.27\%) compared to context (cluster) 2. Because older age, CKD, CVD, and diabetes complications are associated with increased risks of hospitalization, cardiovascular events, and mortality among individuals with T2D \citep{zoungas2014impact, schneider2016diabetes}, we interpret context 1 as a higher-risk group and context 2 as a lower-risk group.

\subsection{Optimal Context-aware Follow-up Policies}
\begin{figure}[H]
  \centering
  \includegraphics[width=1\textwidth]{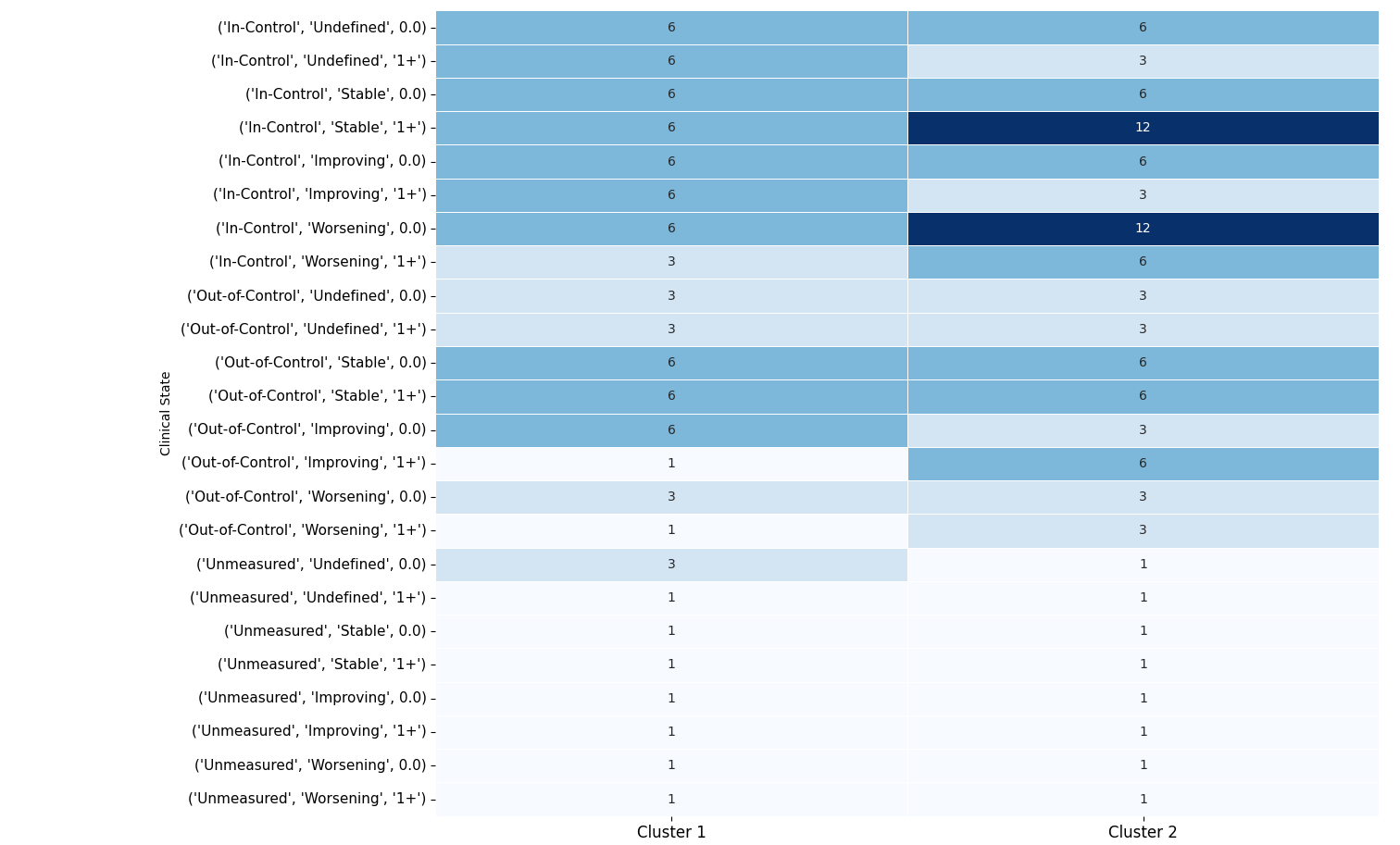}
  \caption{Each row denotes a clinical state s=(glycemic status, short-horizon HbA1c trend, interval hospitalization), with hospitalization coded as 0 (none) or 1+ ($\ge$ 1 admission). Columns correspond to the two patient contexts. Cell values are the optimal follow-up intervals in months \{1,3,6,12\}.}
  \label{fig:policy}
\end{figure}

The CMDP learned context-specific policies, $\pi_z(s)$, for the high-comorbidity (context 1) and lower-comorbidity (context 2) cohorts, as visualized in Figure~\ref{fig:policy}. The policies demonstrate a complex, data-driven stratification of care, with the optimal action depending on both the patient's immediate clinical state and their underlying context.

Three key results emerged from the policies. First, when HbA1c is unmeasured at a given PCP visit, the follow-up policy recommends a 1-month return in both contexts, with only minor deviations when the short-horizon trend is undefined. This result indicates that prolonged periods without measurement elevate the risk of unobserved hyperglycemia, thus the optimal policy prioritizes rapid re-measurement. Second, for ''In-Control'' states, context 2 (the lower-risk group) frequently receives 12-month returns (e.g., for In-Control with Stable or Improving trends), whereas context 1 (older, more comorbid) is capped at 6 months. Conversely, for Out-of-Control states both contexts tighten to 1–3 months, with the shortest interval (often 1 month) triggered by adverse signals such as a worsening HbA1c trend (Worsening) or an interval hospitalization. Third, in a few cases, ``In-Control with an Undefined HbA1c trend'', Context 2 optimal policy recommends a 1-month return while Context 1 optimal policy recommends 6 months. 

Thus, context-aware follow-up policies implement a parsimonious rule set that is easy to operationalize: (i) measure soon when HbA1c is Unmeasured; (ii) lengthen intervals for sustained control, especially in the lower-risk context; and (iii) shorten intervals to 1–3 months when control is poor, trending worse, or when a hospitalization occurred. Crucially, for nearly every clinical state the recommended interval in Context 1 is no longer than in Context 2, indicating that the EHR-derived contexts translate into systematically more intensive follow-up for the higher-risk subpopulation.


\subsection{Comparative Policy Evaluation}
\begin{figure}[htb!]
  \centering
  \includegraphics[width=1\textwidth]{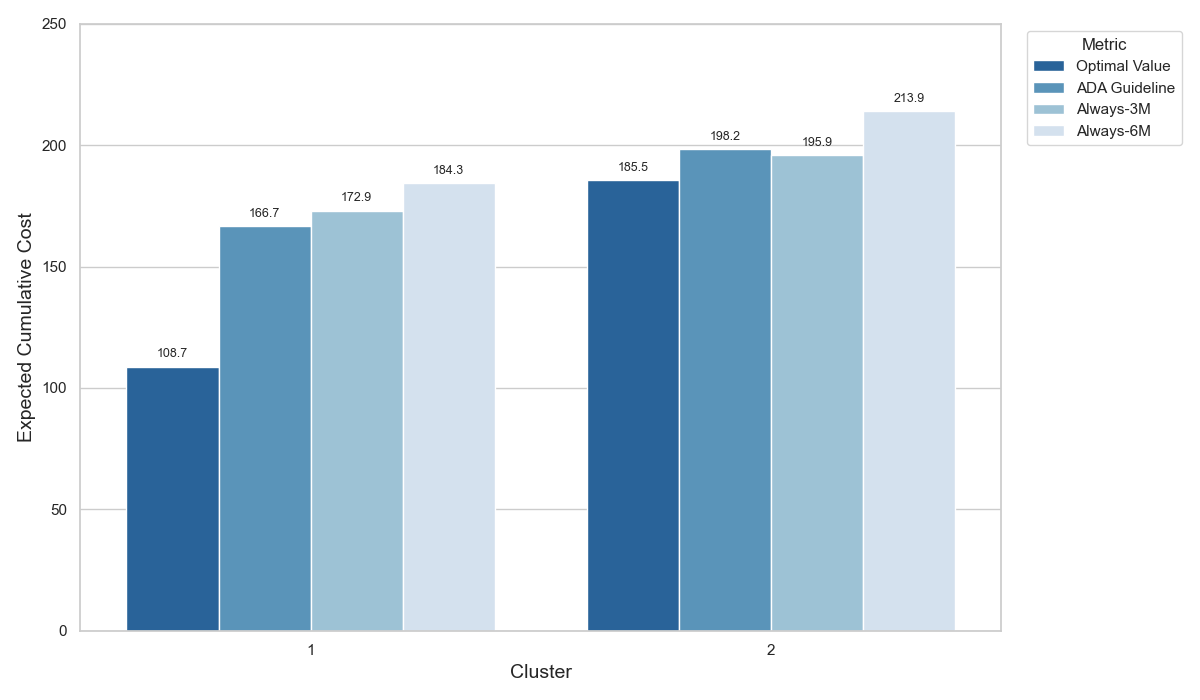}
  \caption{Comparative policy evaluation by patient context. Expected cumulative cost (lower is better) for the learned CMDP policy (Optimal Value) versus three baselines--ADA  rule, fixed 3-month, and fixed 6-month intervals--shown separately for cluster 1 (higher comorbidity) and cluster 2 (lower comorbidity). Numeric labels above bars are absolute costs; the CMDP policy achieves the lowest cost in both clusters.}
  \label{fig:policy_evaluation}
\end{figure}

We benchmarked the learned CMDP policy against three heuristics: (i) fixed ADA guidelines, (ii) a fixed 3-month follow-up policy, and (iii) a fixed 6-month follow-up policy. Figure~\ref{fig:policy_evaluation} shows the expected cumulative cost for each policy, where lower values indicate lower cost. For context 1, the CMDP policy achieved the lowest expected cost (108.7). This was substantially lower than all three baselines: the ADA rule (166.7), the always-3M policy (172.9), and the always-6M policy (184.3). The cost values shown in Figure~\ref{fig:policy_evaluation} are unitless as they combine multiple cost components accumulated between subsequent PCP visits as outlined in Section~\ref{sec:methods}.

For Context 2, the CMDP policy achieved the lowest cost (185.5). This represented a smaller but consistent improvement over the baselines (198.2 for the ADA rule, 195.9 for always-3M, and 213.9 for always-6M). In this context, the fixed 3-month policy was marginally better than the ADA heuristic, but both were inferior to the adaptive CMDP policy.

\subsection{Sensitivity Analysis of Cost Parameters}\label{sec:sensitivity}

\begin{figure}[H]
  \centering
  \includegraphics[width=1\textwidth]{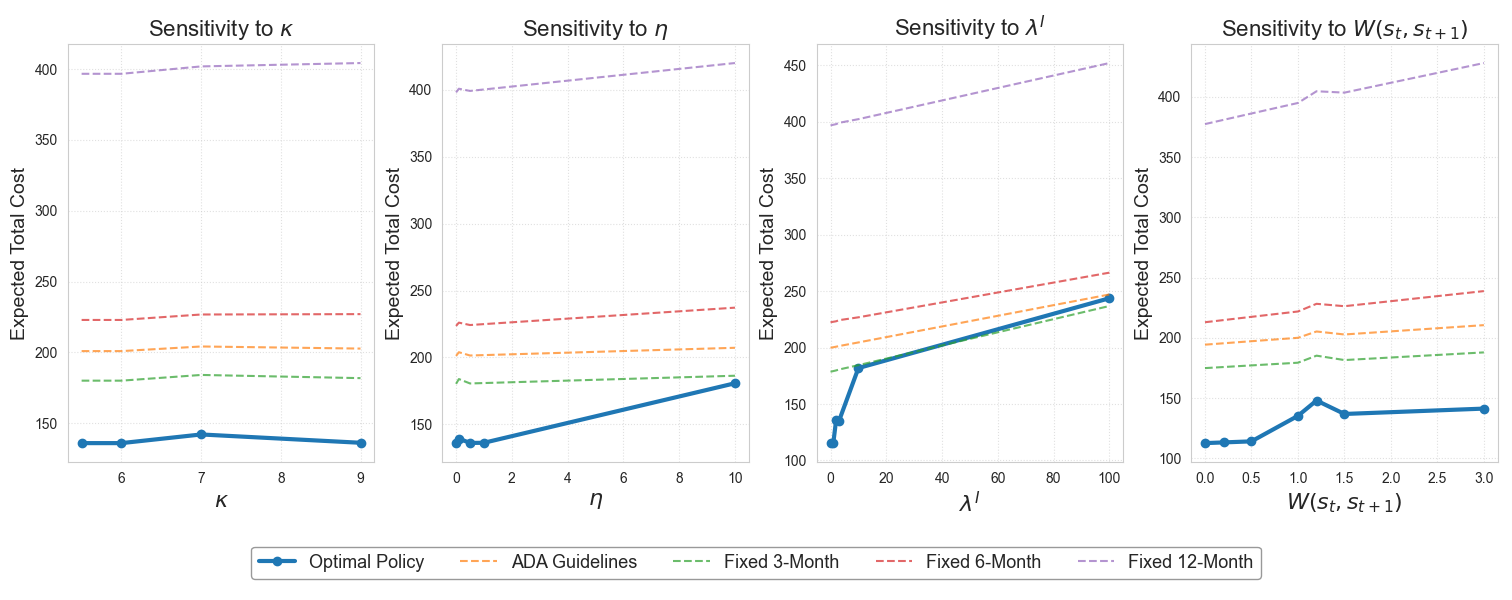}
  \caption{Sensitivity analysis of cost parameters used in CMDP model.}
  \label{fig:comparative_sensitivity_analysis}
\end{figure}

We conducted sensitivity analyses to assess the robustness of the learned CMDP policy to alternative specifications of cost function. We varied four key parameters: (i) the belief HbA1c level $\kappa$ used when glycemic status is unobserved or lost to follow-up in the glycemic burden term; (ii) the baseline uncertainty penalty $\eta$, which scales the cost of acting under outdated or unavailable HbA1c information; (iii) the weight on any inpatient admission between visits, $\lambda_{I}$; and (iv) the state-transition adversity weight $W(s_t, s_{t+1})$ that amplifies glycemic burden for clinically plausible range while holding all other parameters fixed at their baseline values, re-solved the CMDP, and recomputed the expected total cost the test cohort. Figure~\ref{fig:comparative_sensitivity_analysis} compares the CMDP-derived policy with four benchmark strategies (an ADA guideline policy and fixed 3-, 6-, and 12-month follow-up intervals) across these parameters ranges. 


In every scenario, the CMDP policy achieves the lowest expected total cost. The rank ordering of policies remains invariant: the context-aware follow-up policy consistently dominates the guideline and fixed-interval strategies. As $\eta$ and $\lambda_I$ increases, all policies become more costly, which reflects greater penalties for prolonged glycemic uncertainty and hospital admissions; however, the CMDP policy remains clearly separated from comparators, indicating that its advantages is preserved even when decision makers place substantially higher weight on these outcomes. Varying $\kappa$ and $W(s_t, s_{t+1})$ induces only modest shifts in expected cost for all policies, with no evidence of policy cross over. Overall, these analyses show that our main conclusions are robust to reasonable reweighting of the cost components: across a wide range of preferences regarding glycemic burden, measurement uncertainty, and hospitalizations, the CMDP-based policy remains the most efficient strategy in terms of expected clinical burden.

\section{Discussion} \label{sec:discussion}
In this study, we developed and evaluated a Contextual Markov Decision Process (CMDP) framework to personalize return-to-clinic intervals for patients with Type 2 Diabetes (T2D) using de-identified EHR data. By combining trajectory-based PCs with K-Prototypes clustering, we identified clinically meaningful patient contexts and derived context-specific follow-up policies. Across 22,154 patients, the learned policies were interpretable, and aligned with clinical intuition, suggesting that CMDP-based approaches are a feasible pathway toward adaptive, data-driven follow-up decisions in primary care.

Clustering uncovered two distinct and clinically meaningful patient subpopulations. One context consisted of older patients with higher comorbidity burden and more intensive utilization patterns, while the other captured comparatively younger and healthier patients with fewer complications. These contexts emerged from high-dimensional representations of longitudinal trajectories and enabled the CMDP to learn policies that allocate more frequent follow-up to the higher-risk context while avoiding unnecessary visits for lower-risk patients.

Learned policies exhibited clear patterns when expressed over the state space $s_t = (s^H, s^{\Delta H}, s^I)$. For patients in an unmeasured state, both context-aware follow-up policies converged on an intensive 1-month follow-up. This can be interpreted as the lack of recent glycemic data is, in itself, an important signal for follow-up interval decisions. This is aligned with the clinical practice where extended, unobserved periods can increase the risk of undetected hyperglycemia and its associated complications. The model’s  recommendation to re-observe these patients underscores the high cost of uncertainty in managing T2D in primary care setting. For Out-of-Control states, the CMDP favored 1–3 month intervals, with slightly more frequent follow-up when short-term trends indicated worsening control or recent inpatient care. In contrast, for In-Control, stable patients without interval hospitalizations, the model was more willing to extend follow-up to longer intervals, especially in the healthier context. Together, these patterns mirror and refine prevailing recommendations, more frequent assessments in poorly controlled or unstable patients and more relaxed intervals in stable ones, while adding an explicit role for measurement uncertainty and hospitalization history. This asymmetry is plausibly explained by the interaction of (i) the measurement-uncertainty cost (which is larger when trends cannot be computed ) and (ii) context-specific baselines (Context 2 patients are more likely to have limited recent testing despite otherwise good control). By contrast, in Context 1, the combination of adequate recent testing and higher utilization burden can make a 6-month visit optimal when glycemia is currently in control. Finally, the novel cost structure explicitly integrated three clinically relevant dimensions: (i) glycemic burden above a target threshold, (ii) uncertainty induced by infrequent or missing HbA1c measurements, and (iii) the disutility of frequent visits. This cost design forces the CMDP to resolve a genuine trade-off: visiting too often is penalized, but so is failing to measure or control HbA1c promptly, particularly when patients show signs of clinical instability. 

This paper has some limitations that can be addressed in future research. First, the data that used in this study was derived from ten PCPs affiliated with the same healthcare delivery system which may have compromised the generalizability of findings. To address this limitation future should study should include more diverse clinics and study population. Further, the study focuses on in-person primary care follow-up visits and does not model alternative follow-up modalities such as telehealth visits or online portal–based encounters, which are increasingly used in T2D management. Future work should explicitly incorporate these modalities and compare how different follow-up strategies affect visit frequency, laboratory testing, and clinical outcomes. Finally, although the current state representation captures key clinical variables (glycemic status, short-term HbA1c trend, and hospitalizations), it does not include additional information sources such as patient-reported outcomes, medication adherence measures, or continuous glucose monitoring. Incorporating these data sources would allow for more refined context definitions.

\section*{Acknowledgement(s)}
The authors would like to thank the clinicians at MedStar Georgetown University Hospital and data scientists at MedStar Health Research Institute, particularly Laura Schubel for project management support and Sonita S. Bennett for data acquisition and cleaning. The authors are grateful for the funding and support from the Agency of Healthcare Research and Quality (AHRQ). 


\section*{Disclosure Statement}
The authors declare that they have no conflicts of interest.


\section*{Data Availability Statement}
The raw datasets used and/or analyzed during the current study are not publicly available due to data privacy restrictions. Derived and summarized data supporting the findings of this study are available from the corresponding author (MC) on request. 

\section*{Funding}
This work was supported by the Agency of Healthcare Research and Quality (AHRQ) under Grant 1R01HS029792-01.

\bibliographystyle{plainnat}
\bibliography{refs}

\bigskip

\appendix 
\section*{Appendices}
\section{}\label{A} \label{apendixA}
\small 
\begin{longtable}{lp{10cm}}
\caption{List of ICD-10 Codes Representing the Presence of T2D} \label{tab:diag_codes_t2d} \\
\toprule
\textbf{Diagnosis Code} & \textbf{Diagnosis} \\
\midrule
\endfirsthead
\multicolumn{2}{c}%
{\tablename\ \thetable\ -- \textit{Continued from previous page}} \\
\toprule
\textbf{Diagnosis Code} & \textbf{Diagnosis} \\
\midrule
\endhead
\midrule \multicolumn{2}{r}{\textit{Continued on next page}} \\
\endfoot
\bottomrule
\endlastfoot
E11.9   & Type 2 diabetes mellitus without complications \\
E11.40  & Type 2 diabetes mellitus with diabetic neuropathy, unspecified \\
E11.8   & Type 2 diabetes mellitus with unspecified complications \\
E11.65  & Type 2 diabetes mellitus with hyperglycemia \\
E11.319 & Type 2 diabetes mellitus with unspecified diabetic retinopathy without macular edema \\
E11.36	& Type 2 diabetes mellitus with diabetic cataract\\
E11.22	&Type 2 diabetes mellitus with diabetic chronic kidney disease\\
E11.649	&Type 2 diabetes mellitus with hypoglycemia without coma\\
E11.51	&Type 2 diabetes mellitus with diabetic peripheral angiopathy without gangrene\\
E11.21	&Type 2 diabetes mellitus with diabetic nephropathy\\
E11.42	&Type 2 diabetes mellitus with diabetic polyneuropathy\\
E11.52	&Type 2 diabetes mellitus with diabetic peripheral angiopathy with gangrene\\
E11.10	&Type 2 diabetes mellitus with ketoacidosis without coma\\
E11.43	&Type 2 diabetes mellitus with diabetic autonomic (poly)neuropathy\\
E11.69	&Type 2 diabetes mellitus with other specified complication\\
E11.618	&Type 2 diabetes mellitus with other diabetic arthropathy\\
E11.39	&Type 2 diabetes mellitus with other diabetic ophthalmic complication\\
E11.29	&Type 2 diabetes mellitus with other diabetic kidney complication\\
E11.610	&Type 2 diabetes mellitus with diabetic neuropathic arthropathy\\
E11.49	&Type 2 diabetes mellitus with other diabetic neurological complication\\
E11.621	&Type 2 diabetes mellitus with foot ulcer\\
E11.3599 &Type 2 diabetes mellitus with proliferative diabetic retinopathy without macular edema, unspecified eye\\
E11.628	&Type 2 diabetes mellitus with other skin complications\\
E11.00	&Type 2 diabetes mellitus with hyperosmolarity without nonketotic hyperglycemic-hyperosmolar coma (NKHHC)\\
E11.622	&Type 2 diabetes mellitus with other skin ulcer\\
E11.3299 &Type 2 diabetes mellitus with mild nonproliferative diabetic retinopathy without macular edema, unspecified eye\\
E11.3313 &Type 2 diabetes mellitus with moderate nonproliferative diabetic retinopathy with macular edema, bilateral\\
E11.59	&Type 2 diabetes mellitus with other circulatory complications\\
E11.41	&Type 2 diabetes mellitus with diabetic mononeuropathy\\
E11.3291 &Type 2 diabetes mellitus with mild nonproliferative diabetic retinopathy without macular edema, right eye\\
E11.3211 &Type 2 diabetes mellitus with mild nonproliferative diabetic retinopathy with macular edema, right eye\\
E11.3393 &Type 2 diabetes mellitus with moderate nonproliferative diabetic retinopathy without macular edema, bilateral\\
E11.3293 &Type 2 diabetes mellitus with mild nonproliferative diabetic retinopathy without macular edema, bilateral\\
E11.3213 &Type 2 diabetes mellitus with mild nonproliferative diabetic retinopathy with macular edema, bilateral\\
E11.3592 &Type 2 diabetes mellitus with proliferative diabetic retinopathy without macular edema, left eye\\
E11.3552 &Type 2 diabetes mellitus with stable proliferative diabetic retinopathy, left eye\\
E11.3553 &Type 2 diabetes mellitus with stable proliferative diabetic retinopathy, bilateral\\
E11.3593 &Type 2 diabetes mellitus with proliferative diabetic retinopathy without macular edema, bilateral\\
E11.3521 &Type 2 diabetes mellitus with proliferative diabetic retinopathy with traction retinal detachment involving the macula, right eye\\
E11.3591 &Type 2 diabetes mellitus with proliferative diabetic retinopathy without macular edema, right eye\\
E11.3513 &Type 2 diabetes mellitus with proliferative diabetic retinopathy with macular edema, bilateral\\
E11.3533 &Type 2 diabetes mellitus with proliferative diabetic retinopathy with traction retinal detachment not involving the macula, bilateral\\
E11.3523 &Type 2 diabetes mellitus with proliferative diabetic retinopathy with traction retinal detachment involving the macula, bilateral\\
E11.3492 &Type 2 diabetes mellitus with severe nonproliferative diabetic retinopathy without macular edema, left eye\\
E11.3411 &Type 2 diabetes mellitus with severe nonproliferative diabetic retinopathy with macular edema, right eye\\
E11.3391 &Type 2 diabetes mellitus with moderate nonproliferative diabetic retinopathy without macular edema, right eye\\
E11.3212 &Type 2 diabetes mellitus with mild nonproliferative diabetic retinopathy with macular edema, left eye\\
E11.3413 &Type 2 diabetes mellitus with severe nonproliferative diabetic retinopathy with macular edema, bilateral\\
E11.3419 &Type 2 diabetes mellitus with severe nonproliferative diabetic retinopathy with macular edema, unspecified eye\\
E11.3532 &Type 2 diabetes mellitus with proliferative diabetic retinopathy with traction retinal detachment not involving the macula, left eye\\
E11.3542 &Type 2 diabetes mellitus with proliferative diabetic retinopathy with combined traction retinal detachment and rhegmatogenous retinal detachment, left eye\\
E11.3292 &Type 2 diabetes mellitus with mild nonproliferative diabetic retinopathy without macular edema, left eye\\
E11.3219 &Type 2 diabetes mellitus with mild nonproliferative diabetic retinopathy with macular edema, unspecified eye\\
E11.3511 &Type 2 diabetes mellitus with proliferative diabetic retinopathy with macular edema, right eye\\
E11.3493 &Type 2 diabetes mellitus with severe nonproliferative diabetic retinopathy without macular edema, bilateral\\
E11.3412 & Type 2 diabetes mellitus with severe nonproliferative diabetic retinopathy with macular edema, left eye\\
E11.3551 &Type 2 diabetes mellitus with stable proliferative diabetic retinopathy, right eye\\
E11.3519 & Type 2 diabetes mellitus with proliferative diabetic retinopathy with macular edema, unspecified eye\\
E11.311	 &Type 2 diabetes mellitus with unspecified diabetic retinopathy with macular edema\\
E11.3399 &Type 2 diabetes mellitus with moderate nonproliferative diabetic retinopathy without macular edema, unspecified eye\\
E11.3392 &Type 2 diabetes mellitus with moderate nonproliferative diabetic retinopathy without macular edema, left eye\\
E11.3499 &Type 2 diabetes mellitus with severe nonproliferative diabetic retinopathy without macular edema, unspecified eye\\
E11.01	& Type 2 diabetes mellitus with hyperosmolarity with coma\\
E11.3311 & Type 2 diabetes mellitus with moderate nonproliferative diabetic retinopathy with macular edema, right eye\\
\end{longtable}

\small 
\begin{longtable}{lp{10cm}}
\caption{List of ICD-10 Codes Representing the Presence of Chronic Kidney Disease} \label{tab:diag_codes_ckd} \\

\toprule
\textbf{Diagnosis Code} & \textbf{Diagnosis} \\
\midrule
\endfirsthead

\multicolumn{2}{c}%
{\tablename\ \thetable\ -- \textit{Continued from previous page}} \\
\toprule
\textbf{Diagnosis Code} & \textbf{Diagnosis} \\
\midrule
\endhead

\midrule \multicolumn{2}{r}{\textit{Continued on next page}} \\
\endfoot

\bottomrule
\endlastfoot
I12.0	&Hypertensive chronic kidney disease with stage 5 chronic kidney disease or end stage renal disease\\
I12.9	&Hypertensive chronic kidney disease with stage 1 through stage 4 chronic kidney disease, or unspecified chronic kidney disease\\
I13.0	&Hypertensive heart and chronic kidney disease with heart failure and stage 1 through stage 4 chronic kidney disease, or unspecified chronic kidney disease\\
I13.10	&Hypertensive heart and chronic kidney disease without heart failure, with stage 1 through stage 4 chronic kidney disease, or unspecified chronic kidney disease\\
I13.11	&Hypertensive heart and chronic kidney disease without heart failure, with stage 5 chronic kidney disease, or end stage renal disease\\
I13.2	&Hypertensive heart and chronic kidney disease with heart failure and with stage 5 chronic kidney disease, or end stage renal disease\\
N18.1	&Chronic kidney disease, stage 1\\
N18.2	&Chronic kidney disease, stage 2 (mild)\\
N18.3	&Chronic kidney disease, stage 3 (moderate)\\
N18.30	&Chronic kidney disease, stage 3 unspecified\\
N18.31	&Chronic kidney disease, stage 3a\\
N18.32	&Chronic kidney disease, stage 3b\\
N18.4	&Chronic kidney disease, stage 4 (severe)\\
N18.5	&Chronic kidney disease, stage 5\\
N18.9	&Chronic kidney disease, unspecified\\
N19	& Unspecified kidney failure\\
Z94.0	&Kidney transplant status\\
\end{longtable}

\small 
\begin{longtable}{lp{10cm}}
\caption{List of ICD-10 Codes Representing the Presence of Cardiovascular Disease} \label{tab:diag_codes_cvd} \\

\toprule
\textbf{Diagnosis Code} & \textbf{Diagnosis} \\
\midrule
\endfirsthead

\multicolumn{2}{c}%
{\tablename\ \thetable\ -- \textit{Continued from previous page}} \\
\toprule
\textbf{Diagnosis Code} & \textbf{Diagnosis} \\
\midrule
\endhead

\midrule \multicolumn{2}{r}{\textit{Continued on next page}} \\
\endfoot

\bottomrule
\endlastfoot
I25.10	&Atherosclerotic heart disease of native coronary artery without angina pectoris\\
I50.9	&Heart failure, unspecified\\
I63.9	&Cerebral infarction, unspecified\\
I65.23	&Occlusion and stenosis of bilateral carotid arteries\\
I65.29	&Occlusion and stenosis of unspecified carotid artery\\
I67.2	&Cerebral atherosclerosis\\
I67.9	&Cerebrovascular disease, unspecified\\
I73.9	&Peripheral vascular disease, unspecified\\
\end{longtable}

\newpage
\appendix
\section*{Tables}

\begin{table}[H]
    \centering
    \footnotesize
    \setlength{\tabcolsep}{4pt}
    \caption{Baseline characteristics of the study population.}
    \label{tab:characteristics}
    \begin{tabular}{@{} >{\raggedright\arraybackslash}p{3.1cm} l @{\hspace{1em}}
                        >{\raggedright\arraybackslash}p{3.1cm} l @{}}
        \toprule
        \textbf{Characteristic} & \textbf{Count (\%)} & \textbf{Characteristic} & \textbf{Count (\%)} \\
        \cmidrule(lr){1-2}\cmidrule(lr){3-4}

        \textbf{Sex} &                 & \textbf{CKD} & \\
        \quad Female & 12278 (59.94\%) & \quad Yes & 5278 (25.77\%) \\
        \quad Male   & 8202 (40.04\%)  & \quad No  & 15205 (74.23\%) \\
        \quad Unknown& 3 (0.01\%)      &            &                  \\
        \cmidrule(lr){1-2}\cmidrule(lr){3-4}

        \textbf{Race} &                           & \textbf{CVD} & \\
        \quad African American & 13632 (66.55\%) & \quad Yes & 7056  (34.45\%) \\
        \quad White            & 4913 (23.99\%)   & \quad No  & 13427 (65.55\%) \\
        \quad Asian            & 345 (1.68\%)     &           &                 \\
        \quad Other            & 1198 (5.85\%)    &           &                 \\
        \quad Unknown          & 395 (1.93\%)     &           &                 \\
        \cmidrule(lr){1-2}\cmidrule(lr){3-4}

        \textbf{Ethnicity} &                 & \textbf{T2D complication} & \\
        \quad Non-Hispanic & 18824 (91.90\%) & \quad Yes            & 9720 (47.45\%) \\
        \quad Hispanic     & 517 (2.52\%)    & \quad No             & 10763 (52.55\%) \\
        \quad Unknown      & 1142 (5.58\%)   & \quad                &                  \\
        \cmidrule(lr){1-2}\cmidrule(lr){3-4}

        \textbf{Age} &                  & \textbf{Marital status} & \\
        \quad 18--49 & 4476 (21.85\%)   & \quad Single  & 10171 (49.66\%) \\
        \quad 50--59 & 5235 (25.56\%)   & \quad Married & 7823 (38.19\%) \\
        \quad 60--69 & 5794 (28.29\%)   & \quad Widowed & 2420 (11.81\%) \\
        \quad 70+    & 4978 (24.30\%)   & \quad Unknown & 69 (0.34\%) \\
        \bottomrule
    \end{tabular}
    \vspace{4pt}
    \caption*{\footnotesize
      Variables are grouped as demographic (sex, race, ethnicity, age, marital status) and clinical (chronic kidney disease [CKD], cardiovascular disease [CVD], Type~2 diabetes [T2D] complication). Values are counts, with percentages in parentheses.
    }
\end{table}

\newpage
\begin{table}[H]
    \centering
    \footnotesize
    \setlength{\tabcolsep}{4pt}
    \caption{Cluster characteristics (counts with percentages).}
    \label{tab:cluster_outcome}
    \begin{tabular}{@{} >{\raggedright\arraybackslash}p{3.6cm} l l  @{}}
        \toprule
        & \multicolumn{2}{c}{\textbf{Count (\%)}} \\
        \cmidrule(lr){2-3}
        & \textbf{Cluster 1} & \textbf{Cluster 2} \\
        \cmidrule(lr){1-3}

        \textbf{Sex} & &  \\
        \quad Female  & 2278 (64.64\%)  & 10000 (58.97\%) \\
        \quad Male    & 1246 (35.36\%)  & 6956 (41.02\%) \\
        \quad Unknown &  \multicolumn{1}{c}{-} & 3 (0.02\%)     \\
        \cmidrule(lr){1-3}

        \textbf{Age} & & \\
        \quad 18--49 & 441 (12.51\%) & 4035 (23.79\%) \\
        \quad 50--59 & 824 (23.38\%) & 4411 (26.01\%) \\
        \quad 60--69 & 1177 (33.4\%) & 4617 (27.22\%) \\
        \quad 70+    & 1082 (30.7\%) & 3896 (22.97\%) \\
        \cmidrule(lr){1-3}

        \textbf{CKD (ICD)} & &  \\
        \quad CKD    & 1214 (34.45\%)  & 4064 (23.96\%) \\
        \quad No CKD & 2310 (65.55\%)  & 12895 (76.04\%) \\
        \cmidrule(lr){1-3}

        \textbf{CVD (ICD)} & &  \\
        \quad CVD     & 1434 (40.69\%)  & 5622 (33.15\%) \\
        \quad No CVD  & 2090 (59.31\%)  & 11337 (66.85\%) \\
        \cmidrule(lr){1-3}

        \textbf{T2D complication status} & &  \\
        \quad With complication      & 2042 (57.95\%) & 7678 (45.27\%) \\
        \quad Without complication   & 1482 (42.05\%) & 9281 (54.73\%) \\
        \bottomrule
    \end{tabular}
    \vspace{4pt}
    \caption*{\footnotesize
 Abbreviations: CKD = Chronic Kidney Disease; CVD = Cardiovascular Disease.}
\end{table}

\newpage
\appendix
\section*{Figures}

\section*{Figure Captions}

\begin{itemize}
    \item \textbf{Figure 1.} Overview of the methodological approach for deriving context-specific follow-up policies. PC stands for principal component. $n$ represent the number of unique patients in study population. Dynamic variables are observed at each PCP visit at time $t$. After PCA, the dimension is reduced to $k$. $s_t$ represents health state at PCP visit $t$, $a_t$ the follow-up action, $c_t$ cost accumulated between two subsequent PCP visits at times $t_1$ and $t$, and $z$ stands for the context.
    \item \textbf{Figure 2.} Selection of principal components based on cumulative explained variance. The plot shows the cumulative proportion of variance (y-axis) explained by the principal components (x-axis). The dashed vertical line indicates that 83 components were required to capture the pre-specified 95\% (red horizontal line) of the total variance in the standardized feature set.
    \item \textbf{Figure 3.} Each row denotes a clinical state s=(glycemic status, short-horizon HbA1c trend, interval hospitalization), with hospitalization coded as 0 (none) or 1+ ($\ge$ 1 admission). Columns correspond to the two patient contexts. Cell values are the optimal follow-up intervals in months \{1,3,6,12\}.
    \item \textbf{Figure 4.} Comparative policy evaluation by patient context. Expected cumulative cost (lower is better) for the learned CMDP policy (Optimal Value) versus three baselines--ADA  rule, fixed 3-month, and fixed 6-month intervals--shown separately for cluster 1 (higher comorbidity) and cluster 2 (lower comorbidity). Numeric labels above bars are absolute costs; the CMDP policy achieves the lowest cost in both clusters.
    \item \textbf{Figure 5.} Sensitivity analysis of cost parameters used in CMDP model.
\end{itemize}

\end{document}